\documentclass[aps,preprint,pra]{revtex4-1}
\usepackage{graphicx}
\usepackage{amssymb}
\usepackage{amsmath}
\usepackage{color}
\usepackage{enumerate}
\usepackage{bm}
\usepackage[utf8]{inputenc}
\begin{document}
\title{ Disorder-Induced Complex Magnetization Dynamics in Planar Ensembles of Nanoparticles}
\author{Manish Anand}
\email{itsanand121@gmail.com}
\affiliation{Department of Physics, Bihar National College, Patna University, Patna-800004, India.}

\date{\today}
\begin{abstract}
The  magnetic relaxation characteristics are investigated in the two-dimensional ($l^{}_x\times l^{}_y$) assembly of nanoparticles as a function of out-of-plane positional disorder strength $\Delta(\%)$ using numerical simulations. Such defects are redundantly observed in experimentally fabricated nanostructures, resulting in unusual magnetization dynamics.
The magnetization decays exponentially for small and negligible dipolar interaction strength $h^{}_d\leq0.2$. In such a case, the magnetization relaxation does not depend on $\Delta$ and aspect ratio $A^{}_r=l^{}_y/l^{}_x$, as expected. 
In square-like MNPs ensembles and perfectly ordered system ($\Delta(\%)=0$), the magnetization relaxes rapidly with an increase in $h^{}_d$. Consequently, the effective N\'eel relaxation time $\tau^{}_N$ decreases with $h^{}_d$. The dipolar interaction of sufficient strength promotes antiferromagnetic coupling in such a system, resulting in rapid magnetization decay. Remarkably, the out-of-plane disorder instigates the magnetic moment to interact ferromagnetically in the presence of large $h^{}_d$, even in the square-like assembly of MNPs. As a result, magnetization relaxation slows down, resulting in a monotonous increase of $\tau^{}_N$ with an increase in $\Delta$ and $h^{}_d$ in such cases. Notably, there is a prolonged magnetization decay in the highly anisotropic system with large $h^{}_d$. The dipolar interaction induces ferromagnetic coupling along the long axis of the system in such cases. Therefore, the magnetization ceases to relax as a function of time for large $h^{}_d$, irrespective of disorder strength $\Delta(\%)$. 
The present work could provide a concrete theoretical basis to explain the unexpected relaxation behaviour observed in experiments.
These results are also beneficial in digital data storage and spintronics based applications where such nanostructures are extensively used.
\end{abstract}
\maketitle
\section{Introduction}
In recent years, two-dimensional assemblies of magnetic nanoparticles (MNPs) have got tremendous interest because of their unusual magnetic properties~\cite{farhan2013,honolka2009,vsloufova2002,gallina2021structural,leo2018,douarche2003,stariolo2008,verba2020}. They are also instrumental in diverse applications such as data storage, spintronics, magnetic hyperthermia, etc.~\cite{sarella2014,anand2018role,peixoto2020,valdes2021}. The understanding of magnetization dynamics of such systems is of paramount importance for their efficient applications in these physical settings, primarily characterized by a time scale, called the N\'eel relaxation time $\tau^{}_N$~\cite{ilg2020,hergt2009}.
It depends strongly on various system parameters viz particle size, temperature, anisotropy, dipolar interaction, etc.~\cite{iacob2016,bedanta2013}. 


The Arrhenius-N\'eel model successfully describes the relaxation characteristics in the absence of an applied magnetic field and interaction between the nanoparticles~\cite{wegrowe1997,carrey2011}.
However, the dipolar interaction dictates the magnetic relaxation even in dilute MNPs ensembles~\cite{garcia2000,figueiredo2007,morup2010,dormann1988,shtrikman1981}. Figueiredo {\it et al.} investigated the relaxation properties of nanoparticles assembled in triangular lattice~\cite{figueiredo2007}. The dipolar interaction significantly affects the relaxation by reducing the energy barrier in such a case. Morup {\it et al.} and Dormann {\it et al.} observed an enhancement relaxation time due to dipolar interaction~\cite{morup2010,dormann1988}. Shtrikmann and Wohl­farth studied the dipolar interaction dependence of magnetic relaxation~\cite{shtrikman1981}. The dipolar interaction is found to elevate the relaxation time. Because of the long-range and anisotropic nature, the effect of dipolar interaction on the relaxation also depends strongly on the geometrical arrangement and anisotropy axes orientations~\cite{labarta1993,prozorov1999,balcells1997,denisov2003,anand2019,anand2021thermal,gallina2020,farrell2004,anand2021magnetic}. For instance, Denisov {\it et al.} observed the slowing down of magnetic relaxation with dipolar interaction in the two-dimensional MNPs assembly with perpendicular anisotropy~\cite{denisov2003}. In recent work, we investigated relaxation properties in a linear array of nanoparticles as a function of dipolar interaction and anisotropy axis orientation using analytical calculations and kinetic Monte Carlo simulations~\cite{anand2019}. The dipolar interaction increases and decreases the N\'eel relaxation time depending on the angle between the nanoparticles' anisotropy and chain axes. On the other hand, the dipolar interaction always enhances the relaxation time with randomly oriented anisotropy axes~\cite{anand2021thermal}. Gallina {\it et al.} investigated the magnetization dynamics in the two-dimensional assembly of nanoparticles~\cite{gallina2020}. The functional form of the magnetization-decay curve is found to be stretched exponential with small disorder strength. Farrell {\it et al.} prepared two and three-dimensional MNPs ensembles and studied their relaxation characteristics. The relaxation of the three-dimensional system is faster than that of the two-dimensional
arrays~\cite{farrell2004}. Recently, we examined the magnetic relaxation in two-dimensional ordered arrays of nanoparticles using kinetic Monte Carlo (kMC) simulations as a function of dipolar interaction strength and aspect ratio~\cite{anand2021magnetic}. The relaxation characteristics strongly depend on these system parameters.

It is clear from above that the dipolar interaction strongly affects the relaxation characteristics depending on the system sizes, geometrical arrangement of nanoparticles, etc. The two-dimensional ordered systems of MNPs find immense usage in diverse applications such as  data storage, spintronics, etc. The invention of sophisticated techniques to fabricate magnetic nanostructures such as self-assembled MNPs arrays, magnetic thin films, etc., have helped immensely in these aspects~\cite{nolting2000,denisov2001}.  However, such experimentally prepared systems are found to have several types of disorders, which drastically affects their static and dynamic properties~\cite{burmeister1997,rakers1997}. Of all the possible classes of disorders, the out-of-plane positional defect is prevalent. In such a case, all the constituent MNPs in a nanostructure do not lie in a single plane (monolayer) as desired; a few scatter along normal to the sample plane. As a consequence, unexpected relaxation characteristics are observed in such systems. Therefore, it is essential to understand the role of dipolar interaction, defects strength and other system parameters on the magnetization dynamics in such cases. These studies could provide a concrete theoretical basis for explaining and predicting the relaxation behaviours essential for their efficient use in various applications.
Thus motivated, we perform extensive numerical simulations to investigate the effect of out-of-plane disorder strength, dipolar interaction, system size, aspect ratio on the magnetic relaxation
in the two-dimensional MNPs ensembles using kinetic Monte Carlo simulation.

The rest of the article is as follows. We discuss the model and relevant energy terms in Sec.~II. The simulation procedure is also discussed in brief. In Sec.~III, we present and discuss the simulations results. Finally, we provide the summary of the present work in Sec.~IV.  

\section{Theoretical Framework}
\begin{figure}[!htb]
	\centering\includegraphics[scale=0.40]{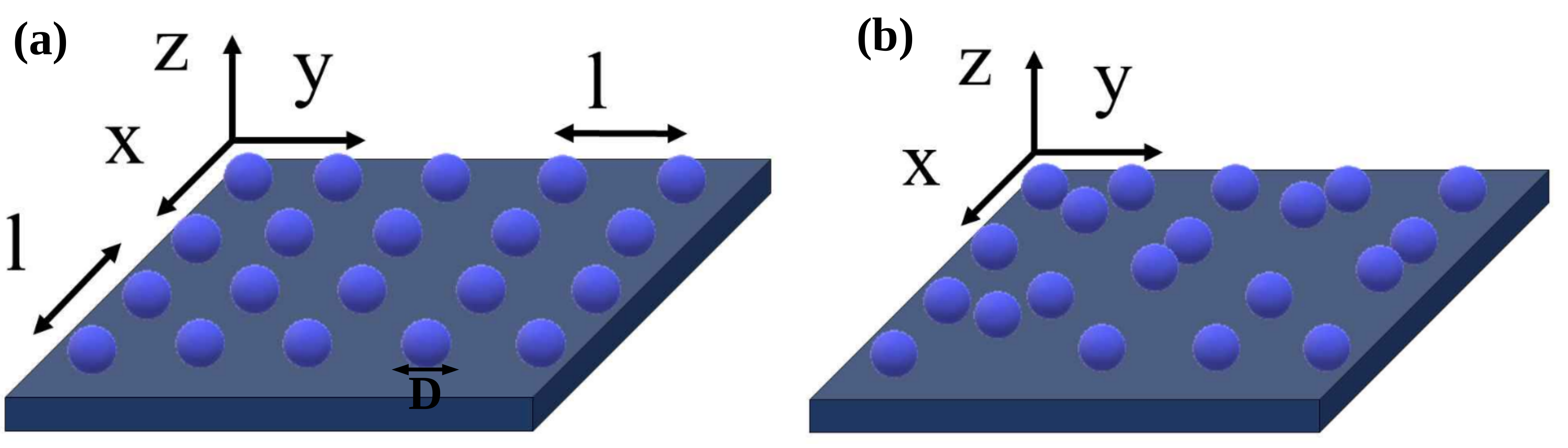}
	\caption{(a) Schematic of the two-dimensional ordered MNPs ensembles. All the nanoparticles are assumed to assemble in the $xy$-plane. $l$ is the lattice constant, and $D$ is the particle diameter. (b) Schematic of the MNPs assembly with the out-of-plane positional disorder. A few nanoparticles are scattered along normal to the sample plane (xy-plane).}
	\label{figure1}
\end{figure}

The nanoparticles are assembled in the two-dimensional lattice ($l^{}_x\times l^{}_y$) in the $xy$-plane, with a few of them  scattered along the $z$-axis, as shown in the schematic Fig.~(\ref{figure1}). The particle has a diameter $D$, and the lattice spacing is $l$, as depicted in Fig.~(\ref{figure1}).
The  magnetocrystalline anisotropy constant associated with each nanoparticle is $K^{}_{\mathrm {eff}}$; $V=\pi D^{3}/6$ is the particle volume. Therefore, the magnetocrystalline energy associated with such a nanoparticle is given by~\cite{anand2016spin,bedanta2008supermagnetism}
\begin{equation}
E=K^{}_\mathrm {eff} V\sin^2\Phi.
\label{barrier}
\end{equation}
Here, the anisotropy axis is assumed to make an angle $\Phi$ with the magnetic moment $\mu=M^{}_sV$; $M^{}_s$ is the saturation magnetization. It is evident from above that the functional form of the anisotropy energy is a symmetric double-well with two minima $E^{0}_1$ (say) and $E^{0}_2$ (say) at $\Phi=0$ and  $\Phi=\pi$, respectively. It also has one energy maximum of strength $E^{0}_3= K^{}_\mathrm {eff} V$ at $\Phi=\pi/2$, also known as the energy barrier. The magnetic moment changes its position from $\Phi=0$ to $\pi$ and vice versa by surmounting the energy barrier in the presence of sufficient thermal fluctuations. The mean time taken by the magnetic moment to perform such flips is termed as the N\'eel relaxation time $\tau^{0}_N$~\cite{anand2019,carrey2011}
\begin{equation}
\tau^{0}_{N}=\tau^{}_o\exp(K^{}_\mathrm {eff}V/k^{}_BT).
\label{Neel}
\end{equation}
$\tau^{}_o$ is the time constant related to attempt frequency $\nu^{}_o\approx10^{10}$ $s^{-1}$ as  $\tau^{}_o=(2\nu_o)^{-1}$. 
$T$ is the temperature, and $k^{}_B$ is the Boltzmann constant. The above relation is applicable to an isolated particle or the extremely dilute assembly of MNPs.

The single-particle energy barrier is primarily modified due to the long-ranged dipolar interaction. The dipolar interaction energy $E^{}_{\mathrm {dip}}$ can be calculated by the following relation~\cite{kechrakos1998,anand2020hysteresis}
\begin{equation}
\label{dipole}
E^{}_{\mathrm {dip}}=\frac{\mu^{}_o}{4\pi l
	^3}\sum_{j,\ j\neq i}\left[ \frac{\vec{\mu_{i}}\cdot\vec{\mu_{j}}-3\left(\vec{\mu_{i}}\cdot\hat{r}_{ij}\right)\left(\vec{\mu_{j}}\cdot\hat{r}_{ij}\right)}{(r^{}_{ij}/l)^3}\right].
\end{equation}
The permeability of free space is $\mu_{o}$; the nanoparticle at $i^{th}$ and $j^{th}$ position has magnetic moment  $\vec{\mu}_{i}$ and $\vec{\mu}_{j}$, respectively. The distance between $i^{th}$ and $j^{th}$ position is $r^{}_{ij}$.
The interaction field $\mu^{}_o\vec{H}^{}_{\mathrm {dip}}$ corresponding to dipolar interaction is given by~\cite{usov2020,tan2014}
\begin{equation}
\mu^{}_{o}\vec{H}^{}_{\mathrm {dip}}=\frac{\mu_{o}}{4\pi l^3}\sum_{j,j\neq i}\frac{3(\vec{\mu}^{}_j \cdot \hat{r}_{ij})\hat{r}^{}_{ij}-\vec{\mu^{}_j} }{(r_{ij}/l)^3}.
\label{dipolar1}
\end{equation}
We model the dipolar interaction strength variation by a control parameter $h^{}_d=D^{3}/l^{3}$~\cite{tan2010}. It correctly captures the physics of interaction strength variation due to the  $1/r^{3}_{ij}$ dependence of $E^{}_{\mathrm {dip}}$ and $\mu^{}_{o}\vec{H}^{}_{\mathrm {dip}}$ [please see Eq.~(\ref{dipole}) and Eq.~(\ref{dipolar1})]. As $h^{}_d=0$ implies $l\to\infty$, it mimics the non-interacting situation. On the other hand, the strongest dipolar interacting case can be modelled by $h^{}_d=1.0$. We can now write the total energy $E$ of the underlying system as~\cite{tan2014,anand2019}
\begin{equation}
E=K^{}_{\mathrm {eff}}V\sum_{i}\sin^2 \Phi^{}_i+\frac{\mu^{}_o\mu^2}{4\pi l^3}\sum_{j,\ j\neq i}\left[ \frac{\hat{\mu_{i}}\cdot\hat{\mu_{j}}-{3\left(\hat{\mu_{i}}\cdot\hat{r}_{ij}\right)\left(\hat{\mu_{j}}\cdot\hat{r}_{ij}\right)}}{(r_{ij}/l)^3}\right]
\end{equation}

The single-particle energy function becomes asymmetric due to the dipolar interaction. Consequently, the energy extrema of the nanoparticle also get altered. Let $E^{}_1$ and $E^{}_2$ be new energy minima and $E^{}_3$ be the modified energy maximum. 
Therefore, the rate $\nu^{}_1$ at which the magnetic moment goes from $E^{}_1$ to $E^{}_2$ by crossing $E^{}_3$ can be evaluated using the following expression~\cite{hanggi1990}

\begin{equation}
\nu^{}_1=\nu^{0}_{1}\exp\bigg(-\frac{E^{}_3-E^{}_1}{k^{}_BT}\bigg)
\end{equation}
Similarly, the jump rate $\nu^{}_2$ to switch its direction from $E^{}_2$ to $E^{}_1$ is given by~\cite{hanggi1990}
\begin{equation}
\nu^{}_2=\nu^{0}_2\exp\bigg(-\frac{E^{}_3-E^{}_2}{k^{}_BT}\bigg),
\end{equation} 
where $\nu^{0}_{1}=\nu^{0}_{2}=\nu^{}_{o}$. 

We have used the kinetic Monte Carlo simulation technique to investigate the out-of-plane disorder effects on the magnetic relaxation in two-dimensional ensembles of nanoparticles. We define a parameter $\Delta(\%)$ to mimic the variation of disorder strength. It is defined as $\Delta(\%)=n^{}_z/n$, where $n$ is the total number of nanoparticles in the system and $n^{}_z$ is the number of particles scattered along the $z$-direction (normal to the system' plane). We also extensively analyze the relaxation mechanism as a function of dipolar interaction strength $h^{}_d$ and aspect ratio $A^{}_r=l^{}_y/l^{}_x$ of the underlying system. The kMC algorithm implemented in the present work is the same as that of Tan {\it et al.} and Anand {\it et al.}~\cite{tan2014,anand2019}. It is also described in greater detail in our recent work~\cite{anand2021tailoring,ANAND2021168461}. Therefore, we do not rewrite it here to avoid repetition. To evaluate the time dependence of magnetization, we apply a huge constant magnetic field $\mu_oH_{\mathrm{max}}$ of strength 20 T along $y$-direction to align all the magnetic moments along the external field direction. After that, the total simulation time is divided into 2000 equal steps, and the applied magnetic field $\mu_oH_{\mathrm{max}}$  is switched off at time $t=0$. We then evaluate the time evolution of magnetization using the kMC algorithm. We fit the so-obtained magnetization decay $M(t)/M^{}_s$ vs $t$ curve with $M(t) = M^{}_s \exp(-t/\tau^{}_N)$ to extract the effective N\'eel relaxation time $\tau^{}_N$.




\section{Simulations Results}
We have taken the following values of parameters for our simulations: $D=8$ nm, $K^{}_{\mathrm {eff}}=13\times10^{3}$ $\mathrm {Jm^{-3}}$, $M^{}_s=4.77\times10^{5}$ $\mathrm {Am^{-1}}$, $n=400$, and $T=300$ K. We have considered five representative values of system sizes $l^{}_x\times l^{}_y=20\times20$, $10\times40$, $4\times100$, $2\times200$, and  $1\times400$. The corresponding value of $A^{}_r(=l^{}_y/l^{}_x)$ is: $1.0$, 4.0, 25.0, 100.0, 400.0, respectively. $\Delta (\%)$ is varied between 5 to 50 to investigate the dependence of the out-of-plane disorder on  magnetic relaxation extensively. We have also varied the dipolar interaction strength $h^{}_d$ from 0 to 1.0. The anisotropy axes are taken as randomly oriented in the three-dimensional space.

\begin{figure}[!htb]
	\centering\includegraphics[scale=0.50]{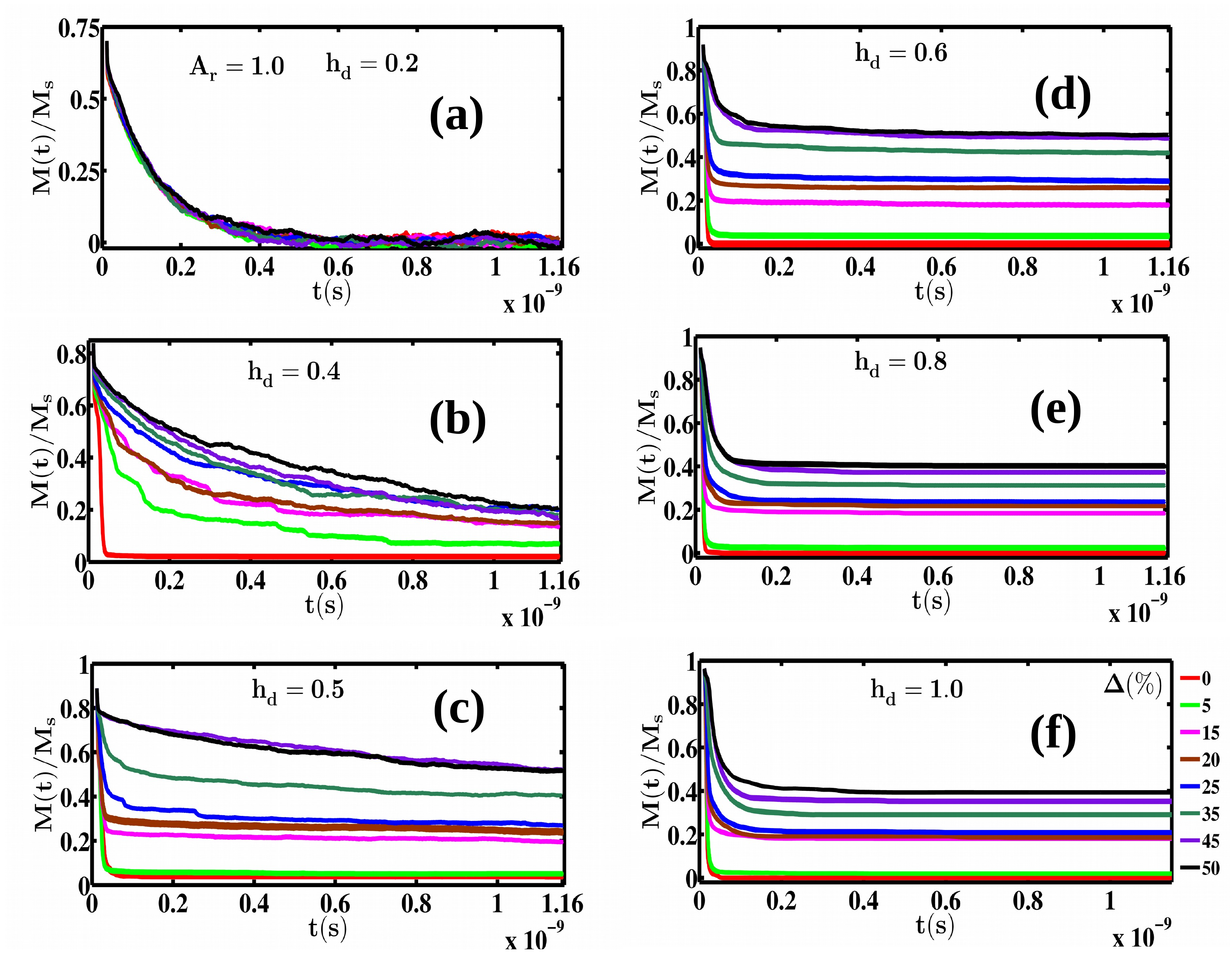}
	\caption{Magnetization-decay curve as a function of out-of-plane disorder $\Delta$ in the square arrays of nanoparticles. We have considered six values of dipolar interaction strength $h^{}_d=0.2$ [a], 0.4 [(b)], 0.5 [(c)], 0.6 [(d)], 0.8[(e)], and 1.0 [(f)]. In the absence of disorder, the magnetization relaxes rapidly with $h^{}_d$. The disorder of enough strength slows down the magentization decay.}
	\label{figure2}
\end{figure}

We first investigate the out-of-plane disorder dependence of magnetic relaxation as a function of dipolar interaction strength in square arrays of MNPs. Fig.~(\ref{figure2}) shows the magnetization-decay $M(t)/M^{}_s$ vs $t$ curve as a function of $h^{}_d$ and $\Delta(\%)$ in the square assembly of nanoparticles ($A^{}_r=1.0$). The out-of-plane disorder strength $\Delta(\%)$ is varied between 0 to 50. We have also considered six representative values of $h^{}_d=0.2$, 0.4, 0.5, 0.6, 0.8, and 1.0. The magnetization decays smoothly in the presence of weak dipolar interaction ($h^{}_d=0.2$). In such a case, the magnetic relaxation is also independent of the disorder strength. In the absence of the disorder ($\Delta(\%)=0$), there is a fastening in magnetization relaxation with an increase in $h^{}_d$. It is because the dipolar interaction of enough strength promotes antiferromagnetic coupling in the square ensembles of MNPs. Consequently, the magnetic moments tend to orient antiferromagnetically in such a case, resulting in a rapid decay of magnetization. Our observation is in qualitative agreement with the work of De'Bell { \it et al.}~\cite{de1997}. They also observed the antiferromagnetic coupling dominance in planar assembly of nanoparticles. Remarkably, the magnetization-relaxation slows down with $\Delta$, provided the dipolar interaction strength $h^{}_d$ is significant. It can be explained by the fact that the nature of the dipolar interaction changes from antiferromagnetic to ferromagnetic in the presence of out-of-plane disorder. In other words, the out-of-plane positional disorder instigates the magnetic moment to interact ferromagnetically for sufficient $h^{}_d$, slowing down magnetization relaxation.

\begin{figure}[!htb]
\centering\includegraphics[scale=0.50]{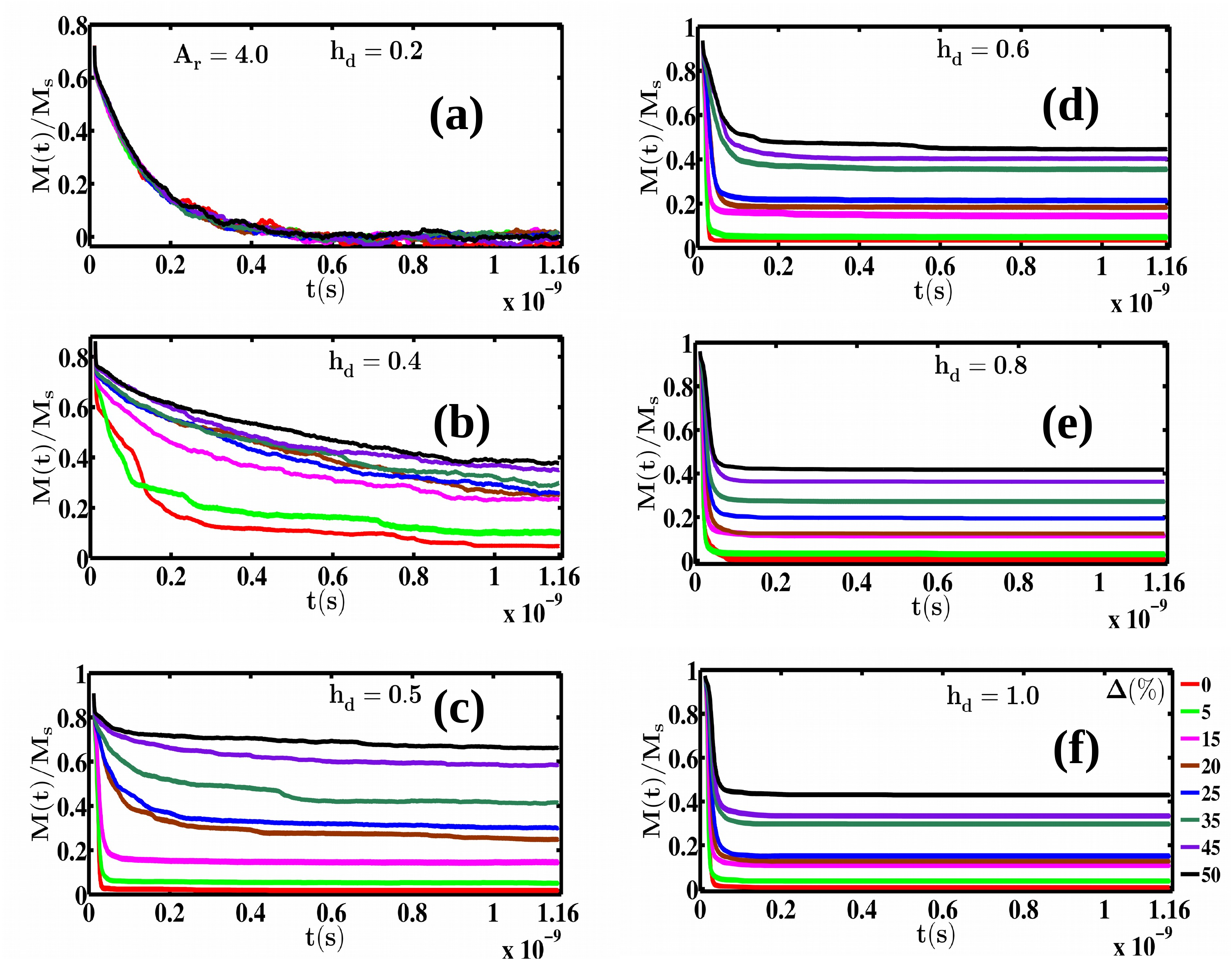}
\caption{Magnetic relaxation as a function of interaction strength $h^{}_d$ and disorder strength $\Delta$ in a rectangular arrays of nanoparticles. Six representative values of $h^{}_d=0.2$ [(a)], 0.4[(b)], 0.5 [(c)], 0.6 [(d)], 0.8 [(e)], and 1.0 [(f)] are considered. We have also varied $\Delta(\%)$ from 0 to 50. There is a slowing down of relaxation with an increase in $h^{}_d$ and $\Delta$.}
\label{figure3}
\end{figure}

We then study the magnetic relaxation as a function of disorder and dipolar interaction strength in the rectangular assembly of nanoparticles. In Fig.~(\ref{figure3}), we plot the time evolution of magnetization-decay $M(t)/M^{}_s$ vs $t$ as a function of $\Delta$ and $h^{}_d$ with $A^{}_r=4.0$. All the other parameters of $h^{}_d$ and $\Delta$ are the same as that of Fig.~(\ref{figure2}). In the presence of weak interaction ($h^{}_d=0.2$), the magnetization-decay characteristic does not depend on $\Delta$  even in the rectangular arrays of nanoparticles. Notably, there is a rapid magnetization-decay with $h^{}_d$ in the case of perfectly ordered MNPs assembly ($\Delta(\%)=0$). It can be attributed to the antiferromagnetic coupling because of dipolar interaction. The rapid decrease 
of magnetization with time is in perfect qualitative agreement with the work of Volkov {\it et al.}~\cite{volkov2008}. Interestingly, the out-of-plane disorder of sufficient strength slows down the magnetization relaxation in the presence of large $h^{}_d$. Such positional defects put the magnetic nanoparticles from antiferromagnetic to ferromagnetic coupling regime in the presence of significant dipolar interaction. These results clearly indicate that we can tune the nature of the dipolar interaction from antiferromagnetic to ferromagnetic and vice versa by just manoeuvring the out-of-plane positional disorder. Therefore, we can also control the magnetization relaxation rate, which could have tremendous implications in data storage and spintronics based applications.  

\begin{figure}[!htb]
\centering\includegraphics[scale=0.50]{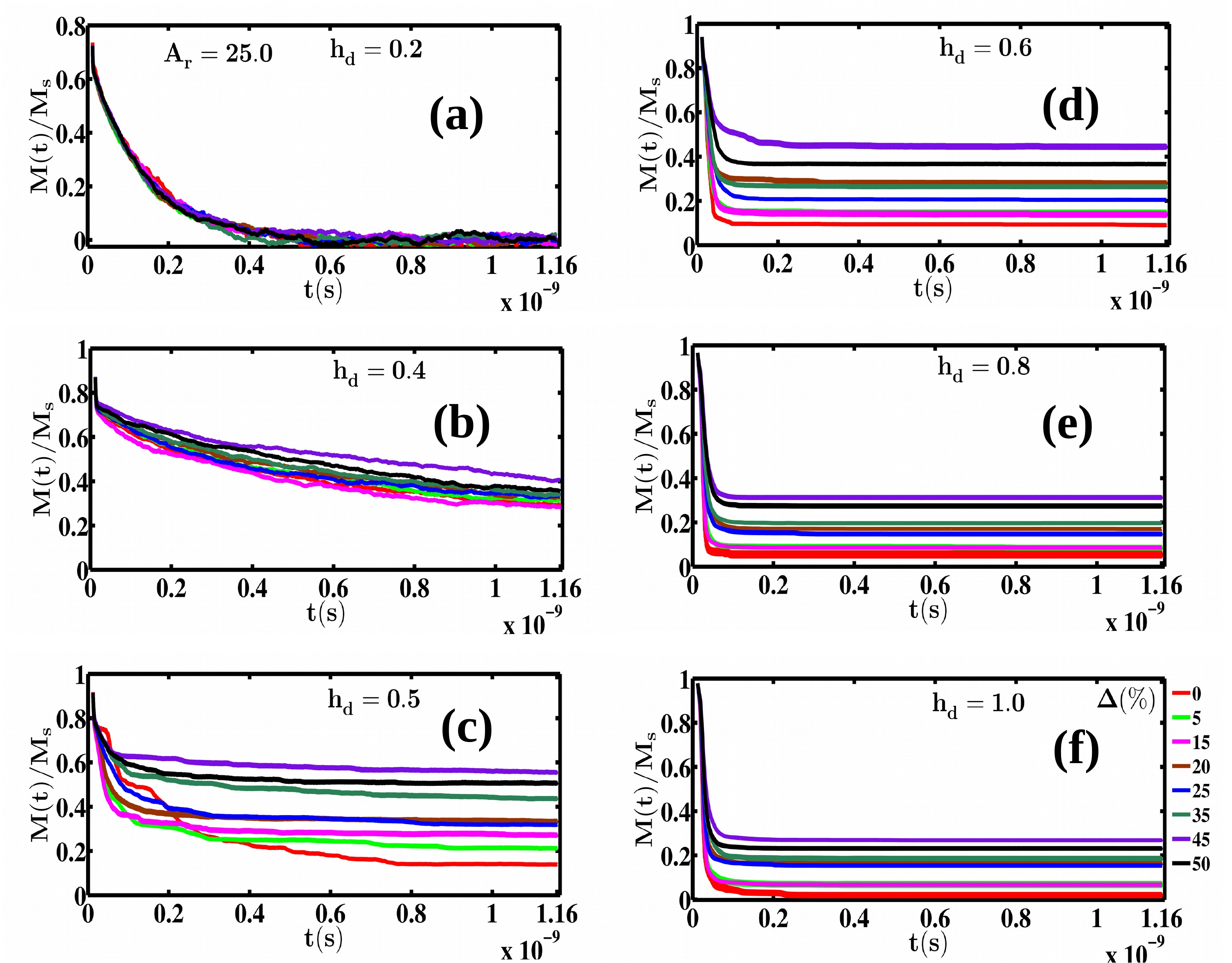}
\caption{The out-of-plane disorder dependence of the magnetic relaxation in a system with aspect ratio $A^{}_r=25.0$. The disorder strength $\Delta(\%)$ is varied between 0 to 50; and six typical values of $h^{}_d=0.2$ [(a)], 0.4[(b)], 0.5[(c)], 0.6[(d)], 0.8[(e)], and 1.0[(f)] are also considered. The nature of the dipolar interaction changes from antiferromagnetic to ferromagnetic in the presence of large $\Delta$, resulting in slowing down of the relaxation.}
\label{figure4}
\end{figure}

\begin{figure}[!htb]
\centering\includegraphics[scale=0.50]{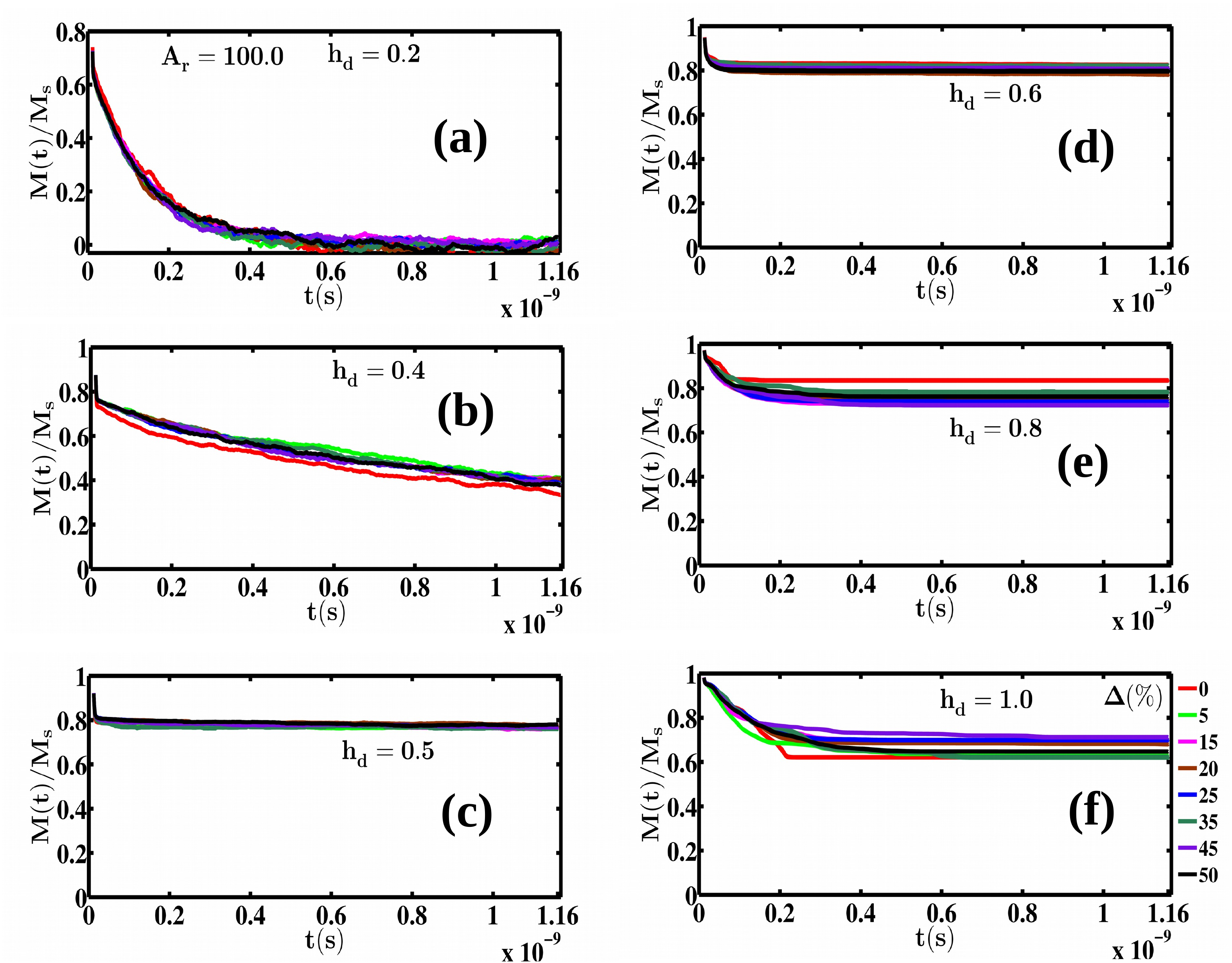}
\caption{Magnetization-decay $M(t)/M^{}_s$ vs $t$ curve as a function of $\Delta$ and $h^{}_d$ with enormous $A^{}_r=100.0$. We have considered six values of $h^{}_d=0.2$ [(a)], 0.4 [(b)], 0.5[(c)], 0.6[(d)], 0.8[(e)], and 1.0 [(f)]. The functional form of the relaxation curve is an exponentially decaying for small $h^{}_d=0.2$. On the other hand, the magnetic ceases to relax even with moderate $h^{}_d=0.4$ because of an enhanced ferromagentic coupling.}
\label{figure5}
\end{figure}

Now, we analyze the relaxation characteristics in a system with enormous aspect ratios to see the anisotropic effect of dipolar interaction. In Fig.~(\ref{figure4}) and Fig.~(\ref{figure5}), we plot the magnetization-decay $M(t)/M^{}_s$ vs $t$ curves as a function of $\Delta$ and $h^{}_d$ with $A^{}_r=25.0$ and  $A^{}_r=100.0$, respectively. The functional form of the magnetization-decay does not depend on $A^{}_r$ and $\Delta$ in the presence of weak dipolar interaction $h^{}_d=0.2$. The magnetization decays rapidly with $h^{}_d$ in a system without positional defects ($\Delta(\%)=0$) and $A^{}_r=25.0$. We can conclude from this observation that the antiferromagnetic coupling is dominant even in a large anisotropic system ($A^{}_r=25.0$). The out-of-plane disorder also changes the nature of dipolar interaction from antiferromagnetic to ferromagnetic in the presence of large $h^{}_d$ and $A^{}_r=25.0$. Consequently, the  magnetic relaxation gets slowed down with $h^{}_d$ and $\Delta$ in such a case. Notably, the magnetization ceases to relax even for moderate dipolar interaction strength $h^{}_d=0.4$ and very large $A^{}_r=100.0$, irrespective of $\Delta$. It is because the dipolar interaction induces ferromagnetic coupling in such a case. Therefore, the magnetic moments tend to interact ferromagnetically, which results in the prolonged decay of magnetization. Even with  significantly large $h^{}_d$, the  magnetization decays extremely slowly, independent of disorder strength.

\begin{figure}[!htb]
\centering\includegraphics[scale=0.50]{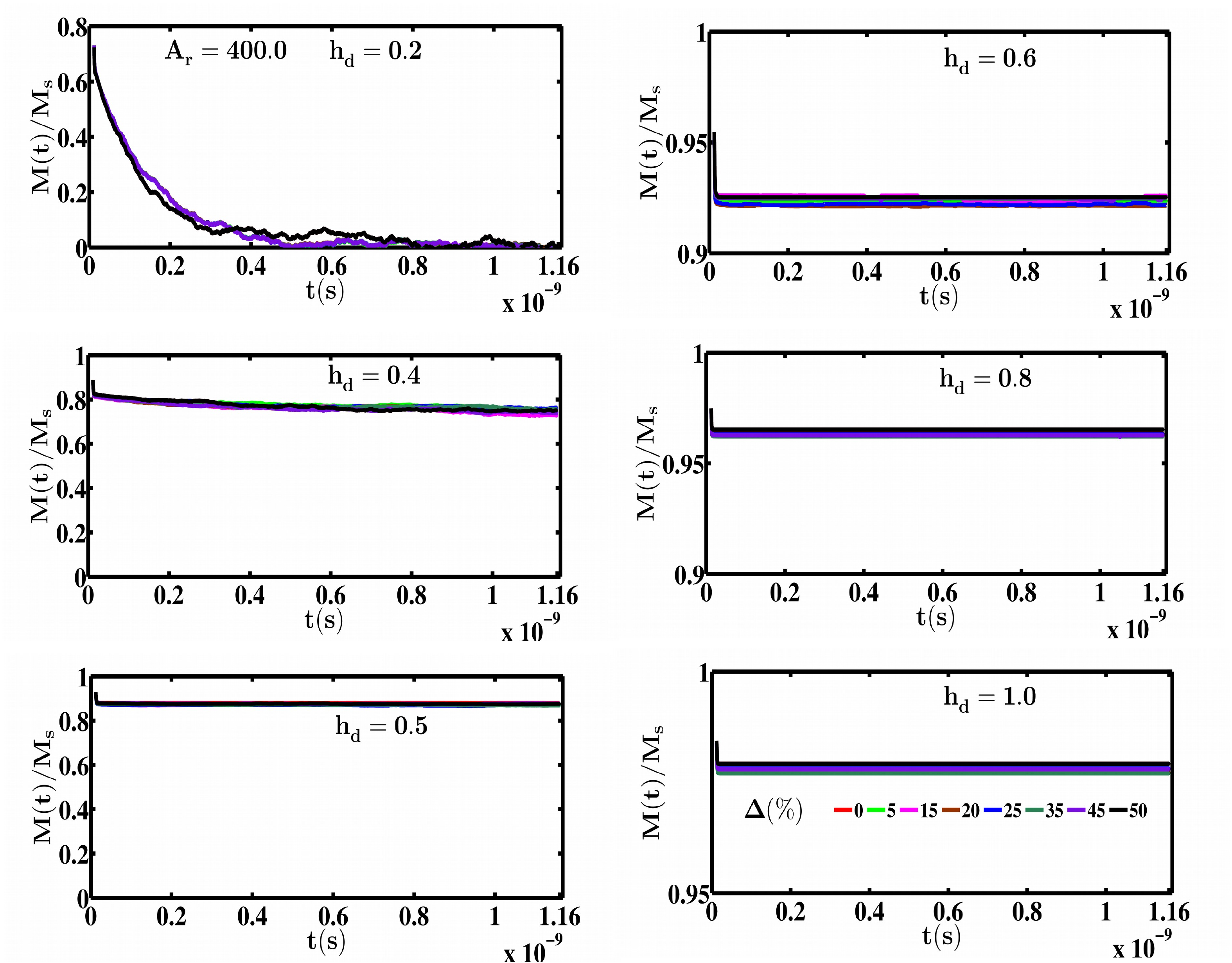}
\caption{The time evolution of the magnetization relaxation in highly anisotropic system $A^{}_r=400.0$. The magnetic relaxation $M(t)/M^{}_s$ vs t curve is shown for six representative values of $h^{}_d=0.2$ [(a)], 0.4[(b)], 0.5[(c)], 0.6[(d)], 0.8[(e)], and 1.0 [(f)]. We have also varied the disorder strength $\Delta(\%)$ from 0 to 50. The ferromagnetic coupling is at its maximum with significant $h^{}_d$. Consequently, the magnetization decays extremely slowly in such cases.}
\label{figure6}
\end{figure}

In Fig.~(\ref{figure6}), we investigate the relaxation characteristics as a function of disorder $\Delta$ and dipolar interaction strength $h^{}_d$ in a highly anisotropic system $A_r=400.0$. There is a smooth decay of magnetization for small $h^{}_d=0.2$, which does not depend on $\Delta$. The magnetization decays exceptionally slowly, even with moderate dipolar interaction strength ($h^{}_d=0.4$). Notably, the disorder does not seem to affect the relaxation behaviour. In such a case, the dipolar interaction creates shape anisotropy along the long axis of the system by inducing ferromagnetic coupling among the constituent nanoparticles. Consequently, the magnetic moments associated with nanoparticles tend to align ferromagnetically along the long axis, resulting in a slowing down of magnetic relaxation. The slowing down of magnetic relaxation with $h^{}_d$ is in qualitative agreement with the work of Osaci {\it et al.}~\cite{osaci2007}. The strength of shape anisotropy or ferromagnetic coupling also gets enhanced with $h^{}_d$ and $\Delta$.  Therefore, the magnetization ceases to relax with an increase in $h^{}_d$, independent of disorder strength $\Delta$.

\begin{figure}[!htb]
\centering\includegraphics[scale=0.50]{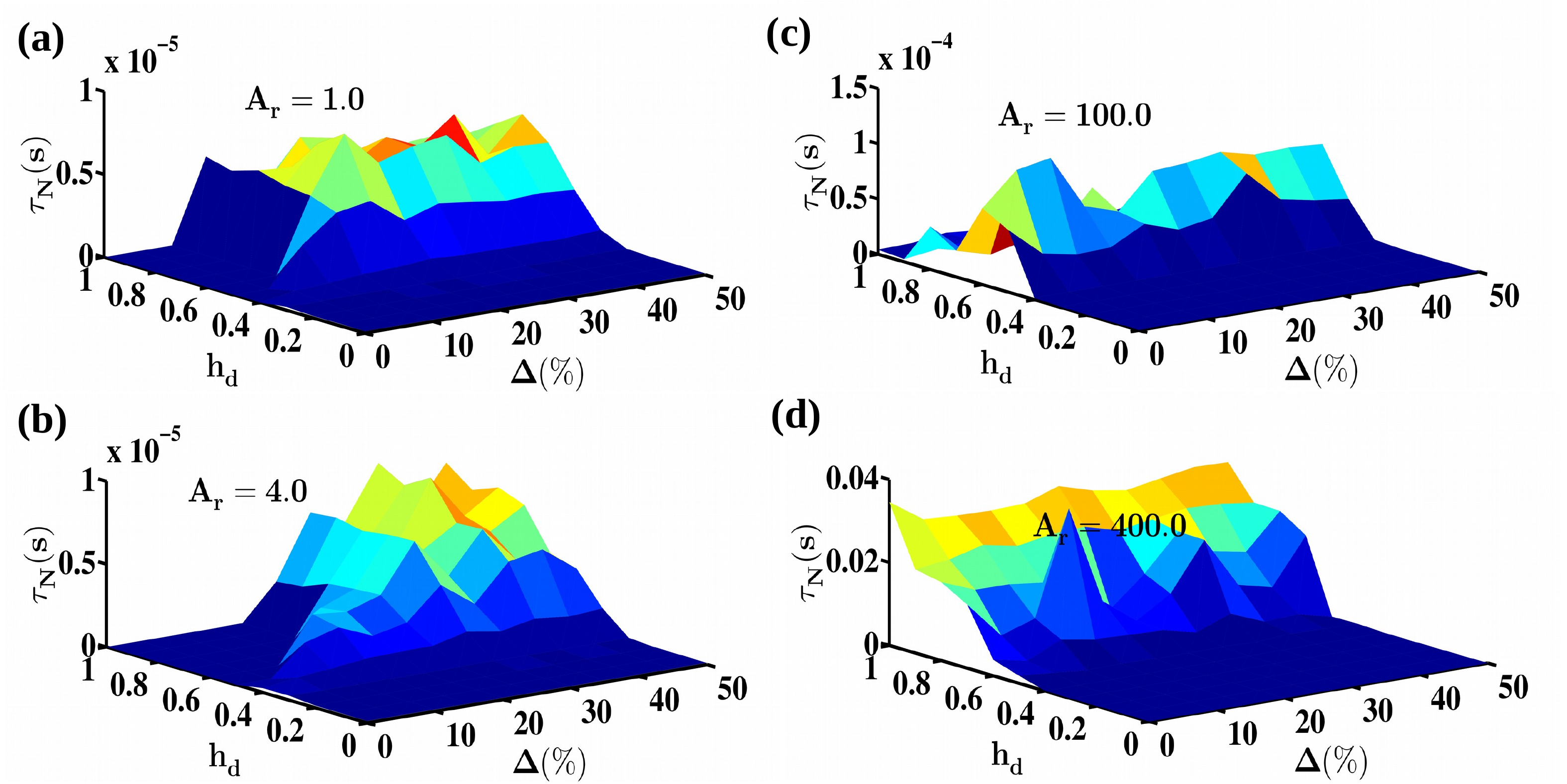}
\caption{The variation of effective N\'eel relaxation time $\tau^{}_N$ as a function of $h^{}_d$ and $\Delta$. We have considered four representative systems with aspect ratio $A^{}_r=1.0$ [(a)], 4.0[(b)], 100.0[(c)], and 400.0[(d)]. In square-like MNPs assemblies ($A^{}_r=1.0$ and 4.0), there is an enhancement in $\tau^{}_N$ with $h^{}_d$ and $\Delta$. Notably,  $\tau^{}_N$ is exceedingly large with huge $A^{}_r=400.0$, and it increases with $h^{}_d$, independent of $\Delta$.} 
\label{figure7}
\end{figure}

 Finally, we study the variation of effective N\'eel relaxation time $\tau^{}_N$ as a function of $h^{}_d$ and $\Delta$ in Fig.~(\ref{figure7}). We have considered four typical values  of aspect ratios $A^{}_r=1.0$, 4.0, 100.0, and 400.0. In the presence of weak interaction strength ($h^{}_d\leq0.2$), $\tau^{}_N$ is significantly small and does not depend on the $\Delta$. In a perfectly ordered system ($\Delta=0$), There is a decrease in $\tau^{}_N$ with $h^{}_d$ in a square-like assembly of nanoparticles ($A^{}_r=4.0$ and $A^{}_r=4.0$). It occurs because the antiferromagnetic coupling gets enhanced with an increase in $h^{}_d$ in such cases. Notably, $\tau^{}_N$ increases with an increase in disorder $\Delta$, provided the  interaction strength $h^{}_d$ is significant. In such a case, the nature of dipolar interaction changes from antiferromagnetic to ferromagnetic. Remarkably, there is always an elevation in $\tau^{}_N$ with an increase in the out-of-plane disorder strength $\Delta$, provided the dipolar interaction strength is significant, and the underlying system is highly anisotropic. It is because the dipolar interaction promotes ferromagnetic coupling in such cases.

\section{Summary and conclusion}
Now, we provide a summary of the present work. We perform computer simulations to study the relaxation characteristics in the two-dimensional ($l^{}_x\times l^{}_y$) assembly of nanoparticles as a function of out-of-plane positional disorder strength $\Delta(\%)$, dipolar interaction strength $h^{}_d$, and aspect ratio $A^{}_r=l^{}_y/l^{}_x$. Such 
positional defects are redundantly observed in experimentally fabricated nanostructures, which drastically affects the relaxation behaviour~\cite{cheng2004magnetic}. Therefore, it is essential to provide a theoretical basis to explain the unexpected magnetic relaxation behaviour due to such defects.  Such studies are also of profound importance because of their immense applications in data storage devices, spintronics, etc. The magnetization-decay $M(t)/M^{}_s$ vs $t$ curve is  exponentially decaying for weak and negligible dipolar interaction ($h^{}_d\leq0.2$). In such a case, the magnetization relaxation does not depend on the disorder strength $\Delta$ and aspect ratio $A^{}_r$, as expected. The simulated value of N\'eel relaxation time also matches with the theoretically calculated [using Eq.~(\ref{Neel})]. In square-like MNPs ensembles and perfectly ordered system ($\Delta(\%)=0$), the magnetization relaxes faster with an increase in $h^{}_d$. Consequently, the effective N\'eel relaxation time $\tau^{}_N$ decreases with $h^{}_d$. It occurs because the dipolar interaction of sufficient strength promotes antiferromagnetic coupling in such a system, resulting in a rapid decay of magnetization. Politi {\it et al.} also observed antiferromagnetic coupling dominance in an assembly of nanoparticles arranged in a triangular lattice~\cite{politi2006}.
Remarkably, the out-of-plane disorder instigates the magnetic moment to interact ferromagnetically in the presence of large $h^{}_d$ even in the square-like assembly of MNPs. In other words, the nature of dipolar interaction changes from antiferromagnetic to ferromagnetic due to the induction of out-of-plane disorder. Consequently, there is a slowing down of magnetization relaxation with $h^{}_d$ and $\Delta$ in such cases. Therefore, there is a monotonous increase of $\tau^{}_N$ with an increase in $\Delta$ and $h^{}_d$ in such cases. Notably, there is a prolonged decay of magnetization in the highly anisotropic system with large $h^{}_d$. The dipolar interaction induces ferromagnetic coupling along the long axis of the system in such cases. Therefore, the magnetization ceases to relax as a function of time for large $h^{}_d$, irrespective of disorder strength $\Delta(\%)$. As a result, $\tau^{}_N$ is also found to be exceedingly large. The slowing down of the magnetization-decay is in perfect agreement with the work of Iglesias {\it et al.}~\cite{iglesias2004}.

In conclusion, we investigated the magnetic relaxation characteristics in the two-dimensional ensembles of nanoparticles as a function of out-of-plane disorder strength. Such positional defects are frequently observed in experimentally designed nanostructures. We also probed the dipolar interaction and aspect ratio effect on the magnetic relaxation behaviour. In the perfectly ordered and square-like assembly, there is a fastening of the magnetic relaxation with increased interaction strength because of the antiferromagnetic coupling dominance. The positional disorder of sufficient strength drives the system from antiferromagnetic to the ferromagnetic regime in the presence of appreciable magnetic interaction, resulting in the slowing down of the magnetization-decay, even in square arrays of nanoparticles. Notably, the magnetization ceases to relax in the highly anisotropic system with significant interaction strength, resulting in negligible magnetization decay. These results are beneficial in explaining the unexpected relaxation behaviour observed in experiments. These observations are also instrumental in diverse technological applications of these nanoparticles assays, such as digital information storage, spintronics based applications, etc.




\section*{DATA AVAILABILITY}
The data that support the findings of this study are available from the corresponding author upon reasonable request.
\bibliography{ref}
\newpage
\newpage
\newpage

\newpage

\newpage

\newpage
\newpage
\end{document}